\documentclass{IEEEtran}
\usepackage{cite}

\usepackage{amsmath,amssymb,amsfonts}
\usepackage{graphicx,hyperref}
\usepackage{textcomp,nicefrac}
\usepackage[switch]{lineno}
\def\BibTeX{{\rm B\kern-.05em{\sc i\kern-.025em b}\kern-.08em
T\kern-.1667em\lower.7ex\hbox{E}\kern-.125emX}}
\begin{document}
\title{Charge Collection Performance of 4H-SiC LGAD}
\author{Sen Zhao, Keqi Wang, Kaibo Xie, Chenxi Fu, Chengwei Wang, Xin Shi and Congcong Wang
\thanks{This work is supported by the state Key Laboratory of Particle Detection and Electronis (202313 and 202312),~Natural Science Foundation of Shandong Province 
Youth Fund (ZR202111120161),~The National Natural Science Foundation of China (12375184,12305207~and 12205321).(Corresponding authors:~Xin Shi; Congcong Wang.)}
\thanks{Sen Zhao is now with the Institute of High Energy Physics, Chinese Academy of Science, Beijing
100049, China,  and also with Hunan Normal University, Hunan 410081, China.}
\thanks{Keqi Wang is now with the Institute of High Energy Physics, Chinese Academy of Science, Beijing
100049, China,  and also with Liaoning University, Liaoning 110136, China.}
\thanks{Kaibo Xie, Chenxi Fu and Chengwei Wang are with the Institute of High Energy Physics, Chinese Academy of Science, Beijing
100049, China.}
\thanks{Xin Shi and Congcong Wang are with
the Institute of High Energy Physics, Chinese Academy of Science, Beijing
100049, China, (e-mail:shixin@ihep.ac.cn;  wangcc@ihep.ac.cn).}}
\maketitle
\begin{abstract}
The 4H-SiC material exhibits good detection performance, but there are still many problems like signal distortion and poor signal quality. The 4H-SiC low gain avalanche detector (LGAD) has been fabricated for the first time to solve these problems, which named SICAR (SIlicon CARbide). The results of electrical characteristics and charge collection performance of the 4H-SiC LGAD are reported. The influence of different metal thicknesses on the leakage current of the device is studied.~By optimizing the fabrication process, the leakage current of the detector is reduced by four orders of magnitude. The experimental results confirm this 4H-SiC LGAD has an obvious gain structure, the gain factor of the SICAR is reported to be about 2 at 150 V. The charge collection efficiency (CCE) of the device was analyzed using $\alpha$ particle incidence of 5.54 MeV, and the CCE is 90\% @100~V. This study provides a novel 4H-SiC LGAD radiation detector for application in the field of high energy particle physics.
\end{abstract}
\begin{IEEEkeywords}
~SiC-LGAD, charge collection efficiency, gain factor
\end{IEEEkeywords}
\section{Introduction}
\label{sec:introduction}

\IEEEPARstart{S}{ilicon} carbide (SiC), as wide band-gap semiconductor, has physical characteristics of wide band-gap, excellent carrier mobility, higher breakdown electric field, higher thermal conductivity and higher saturated drift velocity~\cite{9931717}~compared with silicon~materials. Therefore, it is characterized as higher radiation tolerance~\cite{SELLIN2006479}, lower temperature sensitivity~\cite{Harley-Trochimczyk_2017}, higher breakdown voltage, efficient charge collection and fast time resolution~\cite{Harley-Trochimczyk_2017}. The main common SiC poly-types include 3C-SiC, 4H-SiC, 6H-SiC.~Among them, the 4H-SiC has the widest band-gap (3.26~eV) corresponding to the highest potential radiation hardness~\cite{YANG2023168677}. Radiation detector based on 4H-SiC is suitable for high radiation and extreme temperature environment~\cite{9931717}.\par
In recent years, SiC-based Schottky~\cite{CHAUDHURI2013214,Ruddy_2013,1596541,6418580} and PIN junction~\cite{Torrisi2014DDNF} detectors have been widely used in $\alpha$ particles ~\cite{9931717}, neutron~\cite{WU2014218,LOGIUDICE2007177,WU201372} and photon detection~\cite{Bertuccio2009UltraLN}~\cite{NAVA2003645}. The research shows that the SiC detector has favorable properties such as charge collection efficiency nearly 98\%-100\%~\cite{9931717}, energy resolution about 0.25\%-0.5\% to alpha particles~\cite{Ruddy2006HighresolutionAS}~\cite{Zatko2015HighRA}, fast time resolution about  94~ps~\cite{Yang:2021lli}, radiation resistance nearly three orders of magnitude higher than silicon~detectors~\cite{9217477}~\cite{doi:10.1142/9789812705563_0043}. The 4H-SiC has higher mean ionization energy compared with silicon. As the result, the signal intensity generated by ionization of the same energy in the 4H-SiC detector is smaller than that in the Si detector. In addition, a 4H-SiC detector with a small particle signal will lose some signal instances of a small charge due to the higher energy particle incidence. The collected charge spectrum cannot present a complete Landau distribution, which describes energy loss process of charged particles in a medium.\\

To improve the detection performance of 4H-SiC detector, a 4H-SiC Low Gain Avalanche Detector (LGAD) is proposed to realize carrier avalanche multiplication and amplify the signal~\cite{first_propose}.~In this work, a novel 4H-SiC LGAD has been fabricated for the first time. Its electrical characteristics with different metal thickness and annealing temperatures are systematically evaluated in current-voltage (I-V) and capacitance-voltage (C-V) curves.~Based on the results of experiment,~the charge collection efficiency (CCE) and gain factor of the 4H-SiC LGAD are studied. This work provides a novel SiC detector structure for application in the field of high energy physics.\\

\section{Epitaxial structure and fabrication}

\begin{figure}[h]
\centerline{\includegraphics[width=3in]{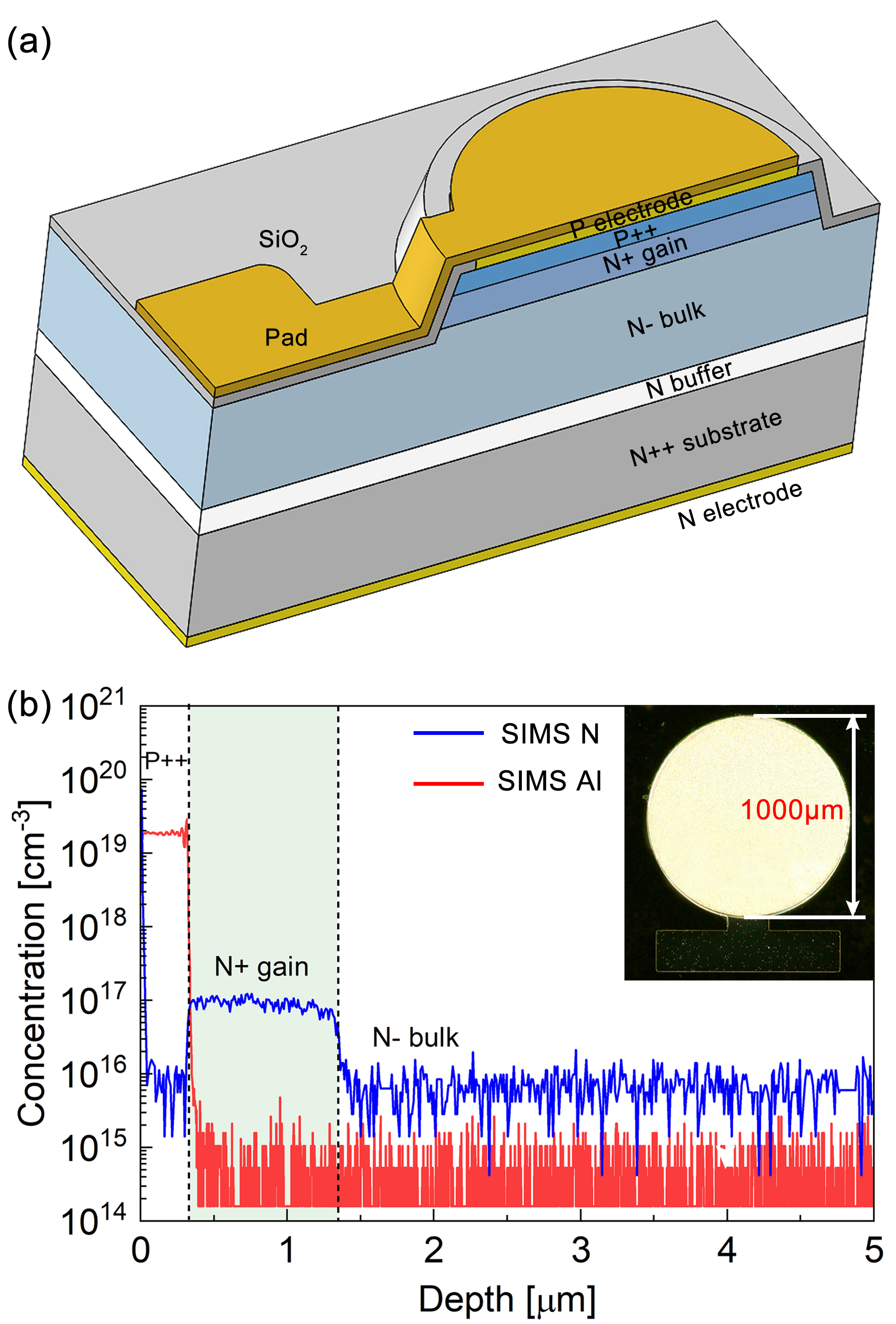}}
\caption{(a)~3D cross-sectional schematic of the 4H-SiC LGAD radiation detector.~(b)~SIMS measurement of doping concentration (characterization depth was 3$\mu$m) and thickness and real 4H-SiC LGAD chip image.}
\label{fig1}
\end{figure}

Epitaxial structure of 4H-SiC LGAD includes P++,~N+ gain,~N- bulk,~N buffer and N++ substrate were first designed~\cite{YANG2023168677}~\cite{P5}.~Secondary ion mass spectrometry (SIMS) was used to characterize the doping concentration and depth of the 4H-SiC LGAD epitaxial structure, and the results were shown in~\figurename~\ref{fig1}~(a). The doping concentration and thickness of epitaxial structure meet our design requirements.~\figurename~\ref{fig1}~(a) shows~physical 4H-SiC LGAD sensor. The structure of 4H-SiC LGAD includes Pad, P electrode, P++, N+ gain, N- bulk, N buffer, N++ substrate, N electrode and SiO$_2$ passivation layer as shown in~\figurename~\ref{fig1}~(b). The fabrication process of 4H-SiC LGAD mainly includes lithography, etching, electron beam evaporation, magnetron sputtering and rapid thermal annealing. Etching N+ gain layer to mitigate premature breakdown caused by the high electric field boundary effect. And etching depth is more than 1.3 $\mu$m to ensure full etching of P++ layer and N+ gain layer. The Ni/Ti/Al metal contact was grown on the top of P++ layer and N++ substrate by using electron evaporating method. SiO$_2$ layer deposited by plasma-enhanced chemical vapor deposition (PECVD) at 350~$^\circ$C. High ohmic contact resistivity is beneficial to improve the electrical performance and charge collection efficiency of 4H-SiC LGAD. Therefore, the ohmic contact electrodes are fabricated using the different metal thickness and annealing temperature to select a 4H-SiC LGAD with the best electrical performance. 

\section{Electrical performance analysis}
\par
To verify the existence of the gain layer in 4H-SiC LGAD, RASER~\cite{P5}~\cite{raser_web} is used to establish the physical models of 4H-SiC LGAD and 4H-SiC PIN based on the carrier Shockley Read-Hall (SRH) composite model. The two detectors have the same structure except for the gain layer. When the bias voltage is high, the barrier capacitance is low, and the diffusion capacitance cannot be ignored.The diffusion capacitance term showed in Eq.~\ref{eq1} needs to be added in C-V simulation,
\begin{equation}
\label{eq1}
C_D=\frac{q(I_n \tau_n+ I_p \tau_p)}{kT} ,
\end{equation}
where, $C_D$ is the diffusion capacitance, $q$ is the charge constant, I$_n$ and I$_p$ are the electron current and hole current, whose value is read from the contact in simulation. $\tau_n$ and $\tau_p$ are the electron lifetime and hole lifetime, whose value is $2.5\times10^{-6}$ s and $0.5\times10^{-6}$ s respectively. $T$ is the temperature, whose value is 300 K. $k$ is the Boltzmann constant.

\begin{figure}[h]
\centerline{\includegraphics[width=3.5in]{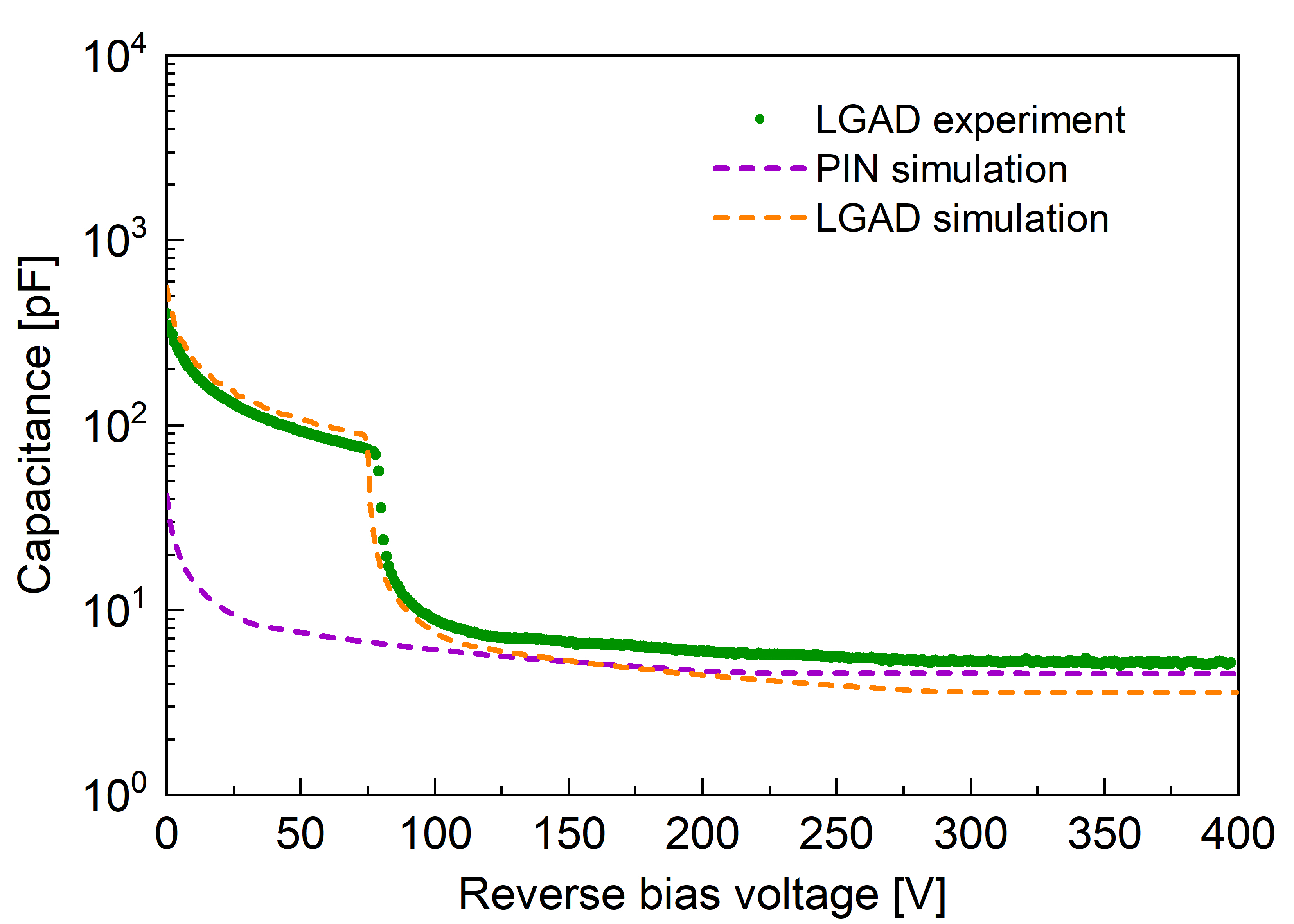}}
\caption{ The C–V results of 4H-SiC LGAD and PIN .}
\label{fig2}
\end{figure}

The  experimental and simulation C-V results are shown in Fig.~\ref{fig2}. The simulated and measured depletion voltages of 4H-SiC LGAD are very close, which indicates that the physical model is correct. Based on this model, the C-V performances of 4H-SiC LGAD and 4H-SiC PIN are analyzed. Compared with the C-V curve of 4H-SiC PIN, it can be seen that the C-V characteristic curve of 4H-SiC LGAD presents an obvious step curve. This proves the PIN detector does not have this feature ~\cite{Yang:2021lli},~and confirms the 4H-SiC LGAD  has an obvious gain structure, which meets our design requirements.

 Contact metal is essential for ohmic contact properties. The Ni/Ti/Al metal is selected for P electrode and N electrode. The Ni only reacts with Si in SiC, C is enriched at the interface of Ni-SiC and inside the Ni$_2$Si, and KIrkendall holes are formed at the interface, roughness of metal interface~\cite{ohmic}.~The existence of Ti layer prevents the enrichment of Ni at the ohmic contact interface and makes the inward diffusion of Ni more uniform, and the flatness of the ohmic contact interface is improved~\cite{Ti/Al/Ni}.~Therefore, the Ni/Ti/Al is selected for ohmic contact  of 4H-SiC. \\

\begin{table}[t]
\caption{\textbf{Metal thickness and annealing conditions for SICARs}}
\label{table}
\setlength{\tabcolsep}{3pt}
\centering
\begin{tabular}{|p{50pt}|p{30pt}|p{30pt}|p{30pt}|p{30pt}|p{30pt}|}
\hline
SICAR samples& sample 1& sample 2& sample 3& sample 4&  sample 5 \\
\hline
Thickness of Ni/Ti/Al (nm)&60/30/80&60/30/80&60/30/80& 60/20/80&50/15/80\\
Temperature ($^\circ$C )&850&950&1050&1050&1050 \\
\hline
\end{tabular}
\label{tab1}
\end{table}
\par
\begin{figure}[h]

\centerline{\includegraphics[width=3.5in]{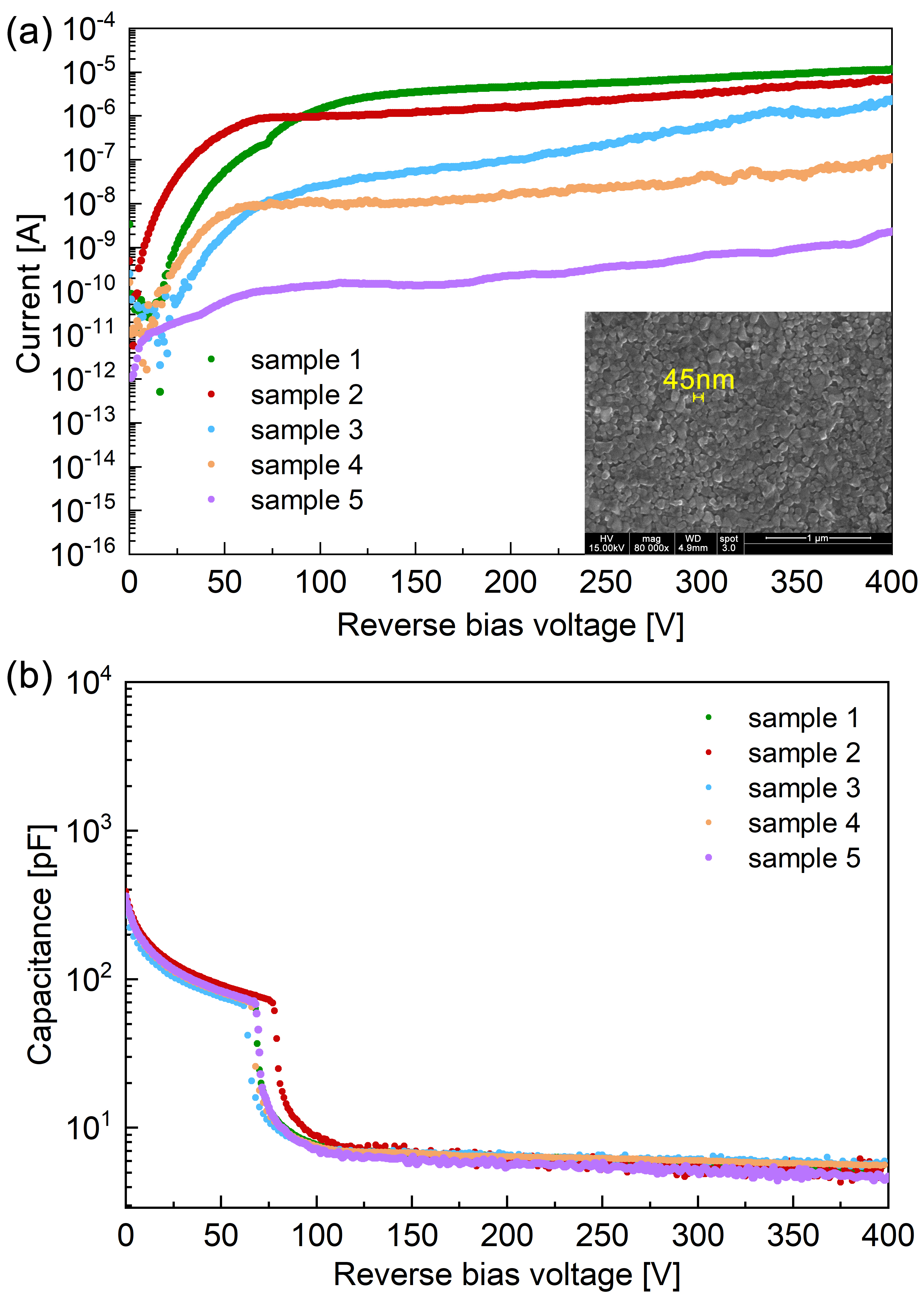}}
\centering
\caption{(a)~I-V characteristics of samples and the SEM image of sample 5.~(b)~C-V characteristics of samples.}
\label{fig3}
\end{figure}

The leakage current characteristics of devices are  shown in the Table~\ref{tab1}~and Fig.~\ref{fig3} (a). From the I-V characteristics of sample 1, sample 2 and sample 3, the leakage current decreases as the annealing temperature increases  when the metal thickness is same. The main reason for this phenomenon is that the annealing temperature of 850~$^\circ$C is too low to rearrange the grains, and more defects on the metal surface. The stress can not be released in time, the metal interface is rough, the ohmic contact characteristics are poor, and the leakage current is large. With the increase of temperature, the internal stress of the metal is released and the metal grains are recrystallized. The metal grain can fill the space between the 4H-SiC lattice. The metal interface film becomes more compact, and the adhesion, stability and ohmic contact performance of the metal film are improved. Therefore, the leakage current of the device is reduced. From the I-V characteristics of sample 3, sample 4 and sample 5,  the leakage current decreases with the decrease of metal thickness when the annealing temperature is the same. The reason for this phenomenon is that if Ni and Ti metals are too thick, there are more lattice defects on the metal surface. The lattice defects form obvious inelastic scattering, and the SiC lattice gaps are overfilled by metal grains to form large volume resistance, which leads to poor ohmic contact effect.~\figurename~\ref{fig3} (a) shows the metal alloy SEM image of sample 5. The SEM image showed that the grain size of the metal is about 45 nm, and the metal film has smooth surface and good compactness. After optimizing the metal thickness and annealing temperature, the leakage of the device is reduced by four orders of magnitude. The leakage current and leakage current density of the sample 5 are 2.4 nA@400 V and 305~nA/cm$^2$@400 V.~The sample 5 has the lowest leakage current. But the leakage current and leakage current density of 4H-SiC LGAD is larger than the same size of PIN in theory~\cite{P5}.~The reason  is that the leakage current generated by thermal excitation and the leakage current due to the increased probability of electron phonon scattering have an amplification effect when passing through the high electric field gain region. All samples can withstand the high voltage of 400~V, which meets the high voltage requirement of particle detector. In summary, suitable annealing temperature and metal thickness can reduce leakage current. 

In order to obtain the effective doping concentration and depletion depth, the C-V characteristics  of all samples are studied as is shown as Fig.~\ref{fig3}~(b). The C-V curves show that all samples present an obvious step curve at a bias voltage of 65 - 80~V,~indicates that the gain layer has been fully depleted at this time. Eq.~\ref{eq2} shows the relationship between effective doping 
concentration (N$_{eff}$) and capacitance.

\begin{equation}
\label{eq2}
N_{eff} = \frac{2}{q\varepsilon A^2} \times \frac{1}{d(1/C^2) / dV} ,
\end{equation}\\
where $A$ is the effective area of the 4H-SiC LGAD, C is the capacitance .
Eq.~\ref{eq3} shows the relationship between depleted depth($W$) and capacitance.

\begin{equation}
\label{eq3}
W=\epsilon \frac{A}{C} .
\end{equation}
\par
The effective doping concentration and depletion depth can be obtained through C-V curves and Eq.~\ref{eq2} and Eq.~\ref{eq3}. The effective doping of the gain region is about $8.6 \times 10^{16}$~cm$^{-3}$ and the depletion depth of the gain layer is about 0.95 $\mu$m. When the voltage reaches 400 V, the depletion depth reaches 29.35 $\mu$m, and the effective doping of the bulk region is about $3.8 \times 10^{14}$ cm$^{-3}$. It can be seen from the experiment that the doping concentration and thickness are within the design range. 

Theory and experiment confirm this 4H-SiC LGAD has an obvious gain structure, the doping concentration and thickness meet our design requirements. The 4H-SiC LGAD with a leakage current of 2.4 nA@400~V were fabricated based on optimizing the metal thickness and annealing temperature.

\section{Charge collection and gain analysis}
\begin{figure}[h]
\centerline{\includegraphics[width=3.5in]{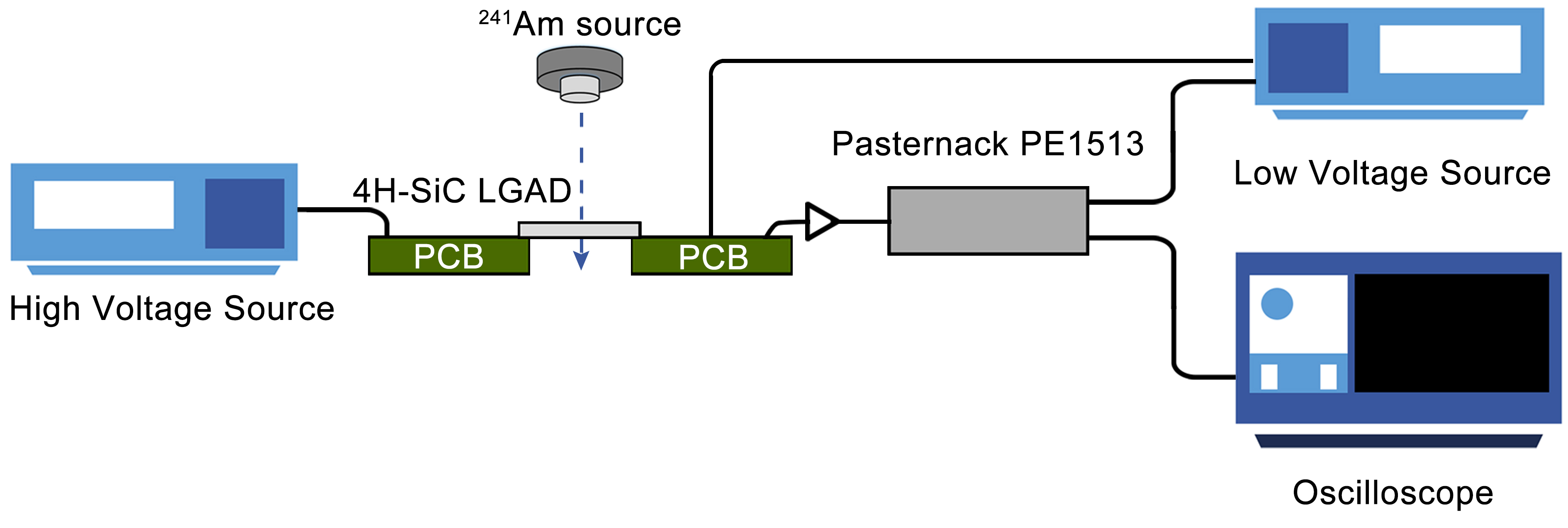}}
\caption{Experimental setup for $\alpha$ particle measurement.}
\label{fig5}
\end{figure}

The charge collection performance of 4H-SiC LGAD with experimental setup shown in the \figurename~\ref{fig5} was investigated. The system includes $^{241}$Am radioactive source, 4H-SiC LGAD, single channel electronic readout board with SiGe amplifier, broadband main amplifier (Model Pasternack 1513), high voltage source (keithley 2470), low voltage source (GPD-3303SGWINSTE) and oscilloscope (DPO-7354C, Tektronix10GHz). The readout board is designed by the University of California, Santa Cruz (UCSC). The board has a 2 mm diameter hole in the middle. The 20 dB broadband main amplifier is placed before the oscilloscope to enhance the SNR. The sampling rate of the oscilloscope is 40 GSa/s and each channel has 20 GSa/s.

The 4H-SiC LGAD is encapsulated on the electronic readout board by using conductive adhesive, and the pad electrode of 4H-SiC LGAD is connected to the readout board. The high voltage source provides the detector with reverse bias. The current signal generated when the particles emitted by the radioactive source pass through the device. The current signal is amplified by an electronic amplifier and converted into a voltage signal. Then the voltage signal is collected by oscilloscope to form pulse waveform. Finally, the charge collection information is obtained by integrating the pulse waveform and converting the current-voltage signal.\\
\begin{figure}[h]
\centerline{\includegraphics[width=3.5in]{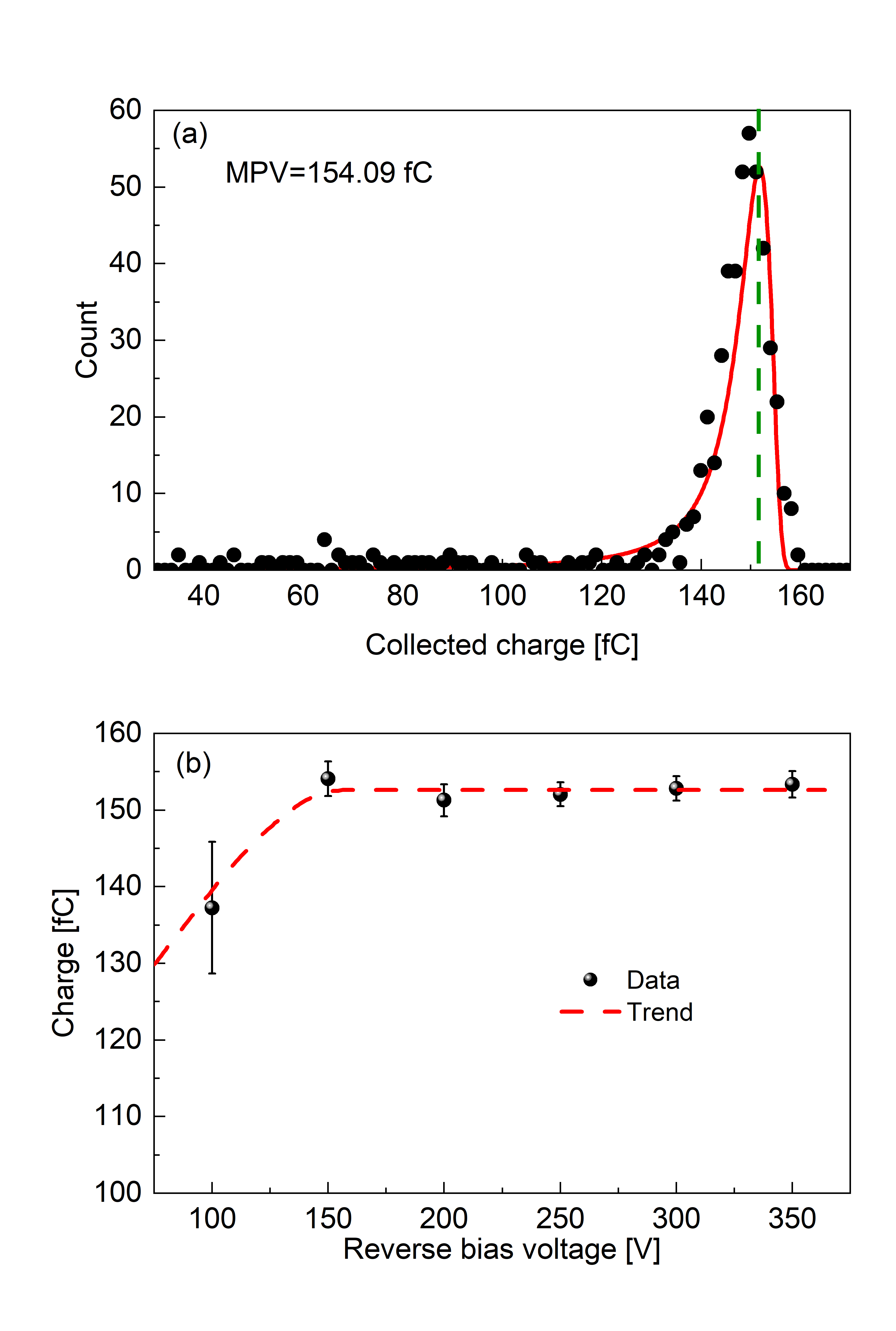}}
\caption{(a)~Charge collection from $\alpha$ particles for 4H-SiC LGAD at 150~V.~(b)~Charge collection of 4H-SiC LGAD at different bias voltages.}
\label{fig6}
\end{figure}

The charge collection distribution of $\alpha$ particles detected by the 4H-SiC LGAD detector at a bias of 150~V is shown as Fig.~\ref{fig6}~(a). The energy loss of $\alpha$ particles through air and metal shows Landau distribution~\cite{LUO2024169110}.~Therefore, a Landau distribution function fitting with the mirror symmetry of the most probable value (MPV) is used to analyze the charge collection. The MPV value is the charge collection. The charge collection of $\alpha$ particles detected by the 4H-SiC LGAD is 152.8~fC@150~V. The same fitting method was used to analyze the charge collection distribution under different bias voltages as shown in Fig.~\ref{fig6}~(b). The charge collection is 137.5 fC@100~V, it shows an increasing trend when the bias voltage is less than 150~V. The charge collection tends to be stable when the bias voltage is 150-350~V. The main reason may be related to the penetration depth of $\alpha$ particles. The $^{241}$Am source decays to release $\alpha$ particles with an energy of 5.54~MeV. It can be fully absorbed by 4H-SiC LGAD devices with 50~$\mu$m bulk layer. According to Eq.~\ref{eq2}, Eq.~\ref{eq3}~and experimental data, when the bias voltage is 100~V and 150~V, the penetration depth of 4H-SiC LGAD are 14~$\mu$m and 17~$\mu$m, respectively. The $\alpha$ particle can no longer ionize more electron hole pairs when the bias voltage is greater than 150~V. The charge collection is stable at around 150~fC. The CCE is defined as 100\% at 150~V, the CCE is 90\% at 100~V.

\begin{figure}[h]
\centerline{\includegraphics[width=3.5in]{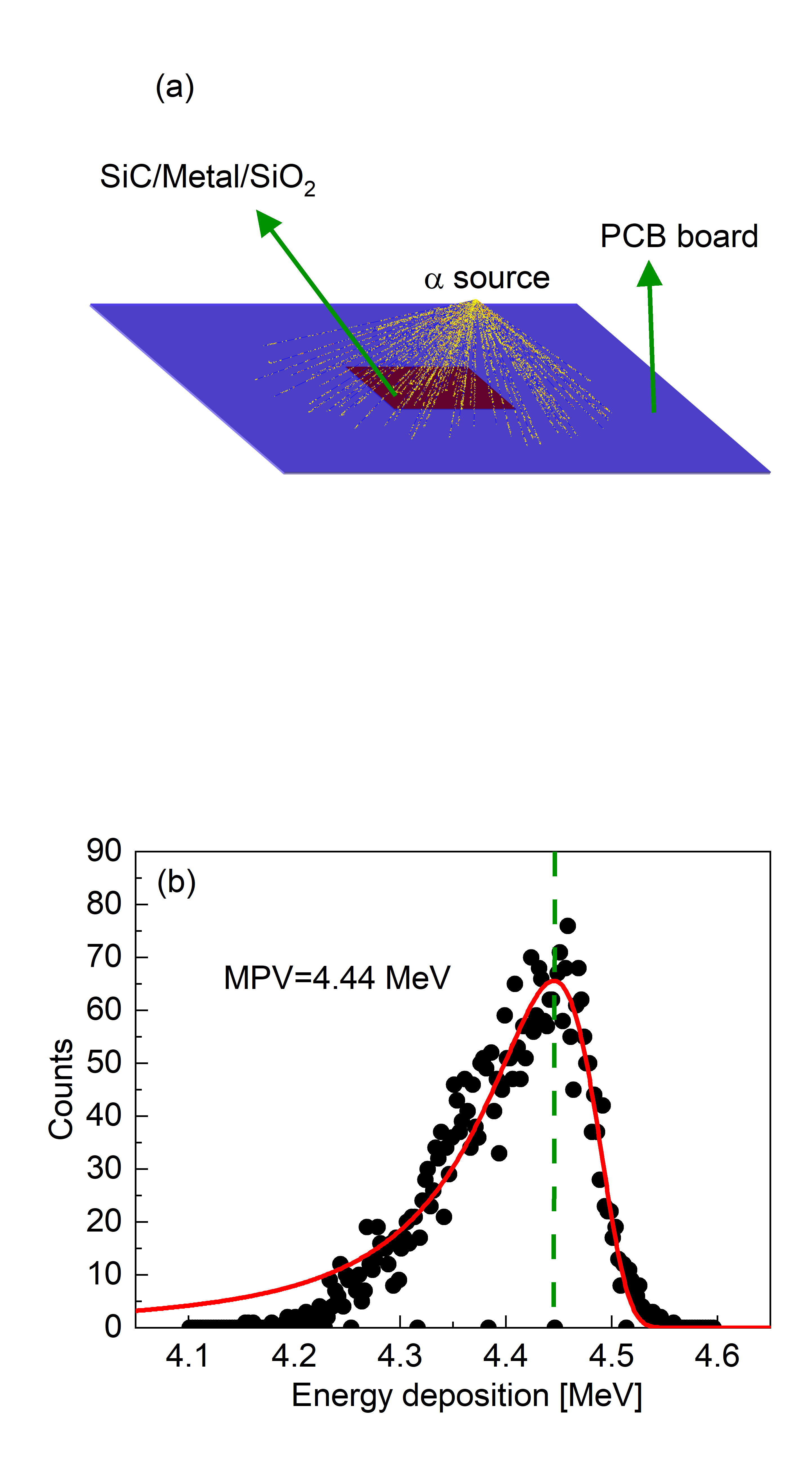}}
\caption{(a)~Schematic diagram of energy deposition simulated by Geant4. (b) The distribution of energy for $\alpha$ deposited in SiC by Geant4 simulation.}
\label{fig7}
\end{figure}

The gain factor of the 4H-SiC LGAD was intensive studied from the perspective of energy deposition. Based on Bethe-Bloch equation, Monte Carlo method and particle transport simulation, the energy deposition of $\alpha$ particles through air, electrode, SiO$_2$ and silicon carbide materials is simulated. Fig.~\ref{fig7} (a) shows the simulation structure and process including $\alpha$ particles (5.54~MeV), 500~nm SiO$_2$, 50~nm Ni, 15~nm Ti, 80~nm Al, 500~nm Al pad and 50~$\mu$m SiC, 0.5~cm distance between $\alpha$ particles and SiC/Metal/SiO$_2$. The simulation parameters are consistent with the parameters of LGAD charge collection experiment except for the grain layer. Fig.~\ref{fig7}~(b) shows the simulated energy deposition with 4.44~MeV. The deposited energy is mainly used for ionization energy loss (IEL) and non-ionization energy loss (NIEL). The IEL is the energy loss of generating electron-hole pairs and exciting electrons to high energy levels. The NIEL is the energy loss of elastic collision with the nucleus of the material. It takes about 8.6~eV energy to ionize one pair of electron-hole in SiC~\cite{Lebedev2004RadiationRO}. Assuming that the energy deposition is all used for the ionization of SiC to generate electron-hole pairs, $5.2 \times 10^5$~of electron-hole pairs can be generated. The charge collection by the detector under the simulated condition is 82.6~fC. Due to the existence of the gain layer, the carrier generated by $\alpha$ particles passing through the gain layer of 4H-SiC LGAD appears avalanche multiplication effect under the action of strong electric field. This leads to additional electron-hole pairs and increased charge collection in 4H-SiC LGAD compared to gainless detectors. Therefore, the charge collection of 4H-SiC LGAD is about 152.8~fC@150~V. In fact, energy deposition cannot be completely used for ionization to generate electron-hole pairs. The gain factor of the 4H-SiC LGAD is around 2 with reverse bias exceeding 150 V.

\section{Conclusion}
The 4H-SiC LGAD radiation detector was fabricated for the first time,which named SICAR, and used in the detection of $\alpha$ particles. The 4H-SiC LGAD radiation detector exhibits consistently low leakage current of 2.4 nA@400~V and 305 nA/cm$^2$@400~V. The charge collection efficiency is about 90\%@100~V. The 4H-SiC LGAD radiation detector is proved to be a good choice in the detection of alpha particles. In the future, we will design and fabricate detectors with a gain factor of 7-9 for detecting position resolution and time resolution. It opens a concrete possibility to use the 4H-SiC LGAD radiation detector for application in the field of high energy particle physics.

\section{Acknowledgement}
This work is supported by the state Key Laboratory of Particle Detection and Electronis (202313 and 202312), Natural Science Foundation of Shandong Province Youth Fund (ZR202111120161), The National Natural Science Foundation of China (12375184,12305207~and 12205321). The authors want to thank Hangrui Shi, Zaiyi Li and RASER team for their help of model simulation and useful discussions.

\bibliographystyle{unsrt}
\bibliography{sample}

\begin{thebibliography}{10}

\bibitem{9931717}
Q.~Yang et~al.
\newblock {High Resolution 4H-SiC p-i-n Radiation Detectors With Low-Voltage
  Operation}.
\newblock {\em IEEE Electron Device Letters}, 43(12):2161--2164, 2022.

\bibitem{SELLIN2006479}
P.~J. Sellin et~al.
\newblock {New materials for radiation hard semiconductor dectectors}.
\newblock {\em Nuclear Instruments and Methods in Physics Research Section A:
  Accelerators, Spectrometers, Detectors and Associated Equipment},
  557(2):479--489, 2006.

\bibitem{Harley-Trochimczyk_2017}
Anna Harley-Trochimczyk et~al.
\newblock {Low-power catalytic gas sensing using highly stable silicon carbide
  microheaters}.
\newblock {\em Journal of Micromechanics and Microengineering}, 27(4):045003, 2
  2017.

\bibitem{YANG2023168677}
T.~Yang et~al.
\newblock {Design and simulation of 4H-SiC low gain avalanche diode}.
\newblock {\em Nuclear Instruments and Methods in Physics Research Section A:
  Accelerators, Spectrometers, Detectors and Associated Equipment},
  1056:168677, 2023.

\bibitem{CHAUDHURI2013214}
S.~K. Chaudhuri et~al.
\newblock {Schottky barrier detectors on 4H-SiC n-type epitaxial layer for
  alpha particles}.
\newblock {\em Nuclear Instruments and Methods in Physics Research Section A:
  Accelerators, Spectrometers, Detectors and Associated Equipment},
  701:214--220, 2013.

\bibitem{Ruddy_2013}
F.~H. Ruddy.
\newblock Silicon carbide radiation detectors: Progress, limitations and future
  directions.
\newblock {\em MRS Proceedings}, 1576:mrss13--1576--ww01--01, 2013.

\bibitem{1596541}
F.~H. Ruddy et~al.
\newblock {High-resolution alpha-particle spectrometry using silicon carbide
  semiconductor detectors}.
\newblock In {\em IEEE Nuclear Science Symposium Conference Record, 2005},
  volume~3, pages 1231--1235, 2005.

\bibitem{6418580}
B.~Zat'ko et~al.
\newblock {Detector of fast neutrons based on silicon carbide epitaxial
  layers}.
\newblock In {\em {The Ninth International Conference on Advanced Semiconductor
  Devices and Mircosystems}}, pages 151--154, 2012.

\bibitem{Torrisi2014DDNF}
L.~Torrisi et~al.
\newblock {D–D nuclear fusion induced by laser-generated plasma at $10^{16}$
  W cm$^{-2}$ intensity}.
\newblock {\em Physica Scripta}, T161, 2014.

\bibitem{WU2014218}
J.~Wu et~al.
\newblock {Effect of neutron irradiation on charge collection efficiency in
  4H-SiC Schottky diode}.
\newblock {\em Nuclear Instruments and Methods in Physics Research Section A:
  Accelerators, Spectrometers, Detectors and Associated Equipment},
  735:218--222, 2014.

\bibitem{LOGIUDICE2007177}
A.~{Lo Giudice} et~al.
\newblock {Performances of 4H-SiC Schottky diodes as neutron detectors}.
\newblock {\em Nuclear Instruments and Methods in Physics Research Section A:
  Accelerators, Spectrometers, Detectors and Associated Equipment},
  583(1):177--180, 2007.

\bibitem{WU201372}
J.~Wu et~al.
\newblock {Feasibility study of a SiC sandwich neutron spectrometer}.
\newblock {\em Nuclear Instruments and Methods in Physics Research Section A:
  Accelerators, Spectrometers, Detectors and Associated Equipment}, 708:72--77,
  2013.

\bibitem{Bertuccio2009UltraLN}
G.~Bertuccio et~al.
\newblock {Ultra Low Noise Epitaxial 4H-SiC X-Ray Detectors}.
\newblock {\em Materials Science Forum}, 615-617:845 -- 848, 2009.

\bibitem{NAVA2003645}
F.~Nava et~al.
\newblock {Radiation tolerance of epitaxial silicon carbide detectors for
  electrons, protons and gamma-rays}.
\newblock {\em Nuclear Instruments and Methods in Physics Research Section A:
  Accelerators, Spectrometers, Detectors and Associated Equipment},
  505(3):645--655, 2003.

\bibitem{Ruddy2006HighresolutionAS}
F.~H. Ruddy et~al.
\newblock {High-resolution alpha-particle spectrometry using 4H silicon carbide
  semiconductor detectors}.
\newblock {\em IEEE Transactions on Nuclear Science}, 53:1713--1718, 2006.

\bibitem{Zatko2015HighRA}
B.~Zat’ko et~al.
\newblock {High resolution alpha particle detectors based on 4H-SiC epitaxial
  layer}.
\newblock {\em Journal of Instrumentation}, 10:C04009 -- C04009, 2015.

\bibitem{Yang:2021lli}
T.~Yang et~al.
\newblock {Time Resolution of the 4H-SiC PIN Detector}.
\newblock {\em Frontiers in Physics}, 10, 2022.

\bibitem{9217477}
J.~M. Rafí et~al.
\newblock {Electron, Neutron, and Proton Irradiation Effects on SiC Radiation
  Detectors}.
\newblock {\em IEEE Transactions on Nuclear Science}, 67(12):2481--2489, 2020.

\bibitem{doi:10.1142/9789812705563_0043}
F.~H. Ruddy et~al.
\newblock {\em FAST NEUTRON SPECTROMETRY USING SILICON CARBIDE DETECTORS},
  pages 347--355.

\bibitem{first_propose}
T.~Yang et~al.
\newblock {{Simulation of the {4H-SiC} Low Gain Avalanche Diode}}.
\newblock {\em arXiv e-prints}, page arXiv:2206.10191, 2022.

\bibitem{P5}
K.~Wang et~al.
\newblock {Design and simulation of a novel 4H-SiC LGAD timing device}.
\newblock {\em Radiation Detection Technology and Methods}, 2023.

\bibitem{raser_web}
X.~Shi.
\newblock Raser.
\newblock \url{https://raser.team/}.
\newblock 2024.5.16.

\bibitem{ohmic}
M.~W. Cole et~al.
\newblock {Improved Ni based composite Ohmic contact to n-SiC for high
  temperature and high power device applications}.
\newblock {\em Journal of Applied Physics - J APPL PHYS}, 88, 05 2001.

\bibitem{Ti/Al/Ni}
R.~Konishi et~al.
\newblock {Development of Ni/Al and Ni/Ti/Al ohmic contact materials for p-type
  4H-SiC}.
\newblock {\em Materials Science and Engineering: B}, 98(3):286--293, 2003.

\bibitem{LUO2024169110}
Y.~Luo et~al.
\newblock {Characterization of the energy response of a LYSO+SiPM detector
  module for E//B NPA using $\alpha$ and hydrogen ions}.
\newblock {\em Nuclear Instruments and Methods in Physics Research Section A:
  Accelerators, Spectrometers, Detectors and Associated Equipment},
  1061:169110, 2024.

\bibitem{Lebedev2004RadiationRO}
A.~Lebedev.
\newblock {Radiation Resistance of SiC and Nuclear-Radiation Detectors Based on
  SiC Films}.
\newblock {\em Semiconductors}, 38:125, 2004.

\end{thebibliography}

\end{document}